\documentclass[a4paper]{jpconf}
\usepackage{graphicx}

\newcommand{\be}{\begin{equation}}
\newcommand{\ee}{\end{equation}}
\newcommand{\ba}{\begin{eqnarray}}
\newcommand{\ea}{\end{eqnarray}}
\newcommand{\adag}{a^\dagger}
\newcommand{\rTr}{{\rm Tr}}

\newcommand{\ma}{\mathcal{A}}

\begin{document}
\title{Recent developments in the shell model Monte Carlo approach to nuclei}

\author{Y. Alhassid$^{1,\ast}$, A. Mukherjee$^{1}$, H. Nakada$^{2}$ and C. \"Ozen$^{3}$}

\address{$^{1}$Center for Theoretical Physics, Sloane Physics Laboratory, Yale University, New Haven, Connecticut 06520, USA\\ $^{2}$Department of Physics, Graduate School of Science, Chiba University, Inage, Chiba 263-8522, Japan\\
$^{3}$Faculty of Engineering and Natural Sciences, Kadir Has University, Istanbul 34083, Turkey}

$^\ast$\ead{yoram.alhassid@yale.edu}

\begin{abstract}
The shell model Monte Carlo (SMMC) approach provides a powerful method for the microscopic calculation of statistical and collective nuclear properties in model spaces that are many orders of magnitude larger than those that can be treated by conventional methods. We discuss recent applications of the method to describe the emergence of collectivity in the framework of the configuration-interaction shell model and the crossover from vibrational to rotational collectivity in families of rare-earth nuclei.  We have calculated state densities of these rare-earth nuclei and find their collective enhancement factors to be correlated with the pairing and shape phase transitions.  We also discuss an accurate method to calculate the ground-state energy of odd-even and odd-odd nuclei, circumventing the sign problem that originates in the projection on an odd number of particles. We have applied this method to calculate pairing gaps in families of isotopes in the iron region.
\end{abstract}

\section{Introduction}

A major challenge in nuclear theory is the derivation of nuclear properties from the underlying effective nuclear interactions. Advances in many-body methods and high-performance computing have enabled significant progress in the application of ab initio methods such as Green's function Monte Carlo~\cite{Pieper2001}, and the no-core shell model~\cite{ncsm-review} and its symplectic extension~\cite{Draayer2008}. However, these methods are limited to light nuclei. Density functional theory (DFT) has made much progress and is unique in providing a global theory of nuclei~\cite{Bender2003} but correlations beyond mean-field theory are often important and require extensions of DFT~\cite{Bender2006}.

The configuration-interaction (CI) shell model approach provides an attractive alternative as it accounts for both shell effects and correlations~\cite{Caurier2005}.
With the help of increasingly more efficient matrix diagonalization algorithms, the CI approach has been successful in describing the low-energy spectroscopy of $fp$-shell nuclei. However, in heavier nuclei conventional diagonalization methods become limited by the combinatorial growth of the dimensionality of the many-particle  space, hindering applications of the conventional CI approach in medium-mass and heavy nuclei.

The shell model Monte Carlo (SMMC) method~\cite{Lang1993,Alhassid1994} has enabled calculations in model spaces that are many orders of magnitude larger than those that can be treated by conventional methods. Monte Carlo methods for fermions are often limited by the so-called sign problem. However, the dominant components~\cite{Zuker1996} of realistic effective nuclear interactions have a good sign and are often sufficient for realistic calculations of statistical and collective properties of nuclei. The smaller, bad sign components can be treated by the method of Ref.~\cite{Alhassid1994}.

Here we discuss recent developments in SMMC, and in particular the application of SMMC in the study of collective and statistical properties of heavy nuclei~\cite{Alhassid2008,Ozen2012}. Another recent progress has been the introduction of a method to calculate the ground-state energy of odd-even and odd-odd nuclei, circumventing a sign problem that originates from the projection on an odd number of particles~\cite{Mukherjee2012}.

\section{The shell model Monte Carlo approach}\label{SMMC}

\subsection{Hubbard-Stratonovich transformation}

The SMMC is based on the Hubbard-Stratonovich (HS) transformation~\cite{HS-trans}, in which the Gibbs operator $e^{-\beta H}$ of a system with an Hamiltonian $H$ at an inverse temperature $\beta=1/T$ is expressed as a functional integral
\be\label{HS}
e^{-\beta H} = \int D[\sigma] G_\sigma U_\sigma
\ee
over external auxiliary fields $\sigma(\tau)$. Here $G_\sigma$ is a Gaussian weight and $U_\sigma$ is a one-body propagator describing non-interacting nucleons moving in external fields $\sigma(\tau)$.

Using the HS transformation, the thermal expectation value of an observable $O$ can be written as
\be \label{observable}
\langle O\rangle = {\Tr \,( O e^{-\beta H})\over  \Tr\, (e^{-\beta H})} = {\int D[\sigma] W_\sigma \Phi_\sigma \langle O \rangle_\sigma
\over \int D[\sigma] W_\sigma \Phi_\sigma} \;,
\ee
where $W_\sigma = G_\sigma |\Tr\, U_\sigma|$ is a positive-definite function and $\Phi_\sigma = \Tr\, U_\sigma/|\Tr\, U_\sigma|$ is known as the Monte Carlo sign function. The quantity $\langle O \rangle_\sigma =
 {\rm Tr} \,( O U_\sigma)/ {\rm Tr}\,U_\sigma$ describes the thermal expectation value of the observable at a given configuration of the auxiliary fields $\sigma$.

 In the SMMC approach $H$ is a rotationally invariant shell-model Hamiltonian defined in a truncated many-particle space spanned by a specified set of $N_s$ single-particle orbitals (this set can be different for protons and for neutrons).  The quantities in the integrands of Eq.~(\ref{observable}) can be expressed in terms of the $N_s \times N_s$ matrix ${\bf U}_\sigma$ representing $U_\sigma$ in the single-particle space. For example, the grand-canonical trace of $U_\sigma$ is given by
${\rm Tr}\; U_\sigma = \det ( {\bf 1} + {\bf U}_\sigma)$.

In the finite nucleus it is important to use the canonical ensemble with fixed number of protons and neutrons. This is done using particle-number projection, which for a finite number of orbitals can be expressed as a discrete Fourier sum. For example, the trace of $U_\sigma$ at fixed particle number $A$ is given by
\begin{eqnarray}\label{canonical}
\Tr_A U_\sigma =\frac{e^{-\beta\mu  A}}{N_s}
\sum_{m=1}^{N_s} e^{-i\phi_m A}\,\det \left({\bf 1}+e^{i\phi_m}e^{\beta\mu}{\bf U}_\sigma\right)
\;,
\end{eqnarray}
where $\hat A$ is the particle-number operator, $\phi_m=2\pi m/N_s \;\; (m=1,\ldots,N_s)$ are quadrature points and $\mu$ is a real chemical potential [required for the numerical stabilization of the sum in (\ref{canonical})].

 \subsection{Monte Carlo method}

 The imaginary-time interval $(0,\beta)$ is discretized in steps of $\Delta \beta$. The integration over the auxiliary fields is of very high dimension and is carried out by a Monte Carlo method. The integration over each $\sigma$ field at a given time is approximated by a three-point quadrature formula and importance sampling is used to choose $M$ uncorrelated configurations $\sigma_k$ according to the positive-definite weight function $W_\sigma$. The expectation value in (\ref{observable}) is then estimated from
 \be\label{MC-average}
\langle O\rangle \approx  {\sum_k
  \langle  O \rangle_{\sigma_k} \Phi_{\sigma_k} \over \sum_k \Phi_{\sigma_k}} \;.
\ee

\section{Collectivity in the configuration-interaction shell model approach}\label{collectivity}

Heavy nuclei are known to exhibit collectivity in their low-energy spectra and electromagnetic transition rates.  Phenomenological nuclear structure models describe well the various types of collectivity such as vibrational collectivity in spherical nuclei and rotational collectivity in deformed nuclei. However, much less is understood about the microscopic emergence of collectivity in the framework of a CI shell model approach.  A shell model treatment of heavy deformed nuclei requires very large model spaces and that are accessible by the SMMC method.

We have extended the SMMC method to heavy nuclei using a proton-neutron formalism in which the protons and neutrons can occupy different shells~\cite{Alhassid2008}. In particular, we have applied the method to rare-earth nuclei using the $50-82$ major shell plus the $1f_{7/2}$ orbital for protons and the $80-126$ major shell plus the $0h_{11/2}$ and $1g_{9/2}$ orbitals for neutrons.

The SMMC method enables us to calculate thermal and ground-state observables in very large model spaces, but it does not provide detailed spectroscopic information. Since nuclear collectivity is often identified by its spectroscopic characteristics, it is necessary to find a thermal observable that is sensitive to the specific type of collectivity. We have identified such an observable in $\langle {\bf J}^2\rangle_T$, where ${\bf J}$ is the total angular momentum of the nucleus~\cite{Alhassid2008,Ozen2012}. For an even-even nucleus and at sufficiently low temperatures, this observable is given by
\begin{eqnarray}\label{J2-theory}
\langle \mathbf{J}^2 \rangle_T \approx
 \left\{ \begin{array}{cc}
 30 { e^{-E_{2^+}/T} \over \left(1-e^{- E_{2^+}/T}\right)^2} &{\rm vibrational\; band}  \\
 \frac{6}{E_{2^+}} T & {\rm rotational \;band}
 \end{array} \right.
\end{eqnarray}
where $E_{2^+}$ is the excitation energy of the first $2^+$ level.

\subsection{Rotational collectivity in heavy deformed nuclei}
\begin{figure}[h!]
    \includegraphics[clip,angle=0,width=\textwidth]{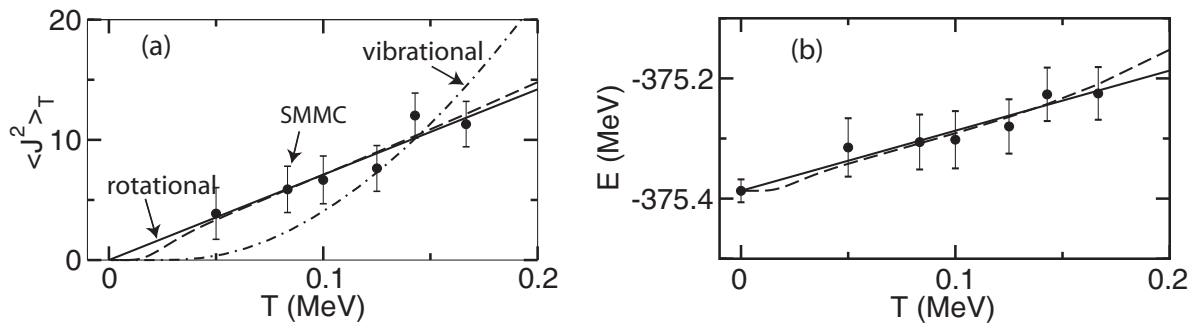}
    \caption{ (a) $\langle {\bf J}^2\rangle_T$ and (b) thermal energy $E(T)$ versus $T$ for $^{162}$Dy.  The SMMC results (solid circles with error bars) are compared in (a) with fits to the rotational model (solid line) and vibrational model (dotted-dashed line) [see Eq.~(\ref{J2-theory})], and in (b) with the rotational model (solid line). The dashed lines are obtained from the lowest five experimental rotational bands.  Adapted from Ref.~\cite{Alhassid2008}.}
    \label{low-T}
\end{figure}
An important issue is whether we can describe the rotational collectivity of heavy deformed nuclei in the framework of a truncated spherical shell model space. In Ref.~\cite{Alhassid2008} we studied a strongly deformed rare-earth nucleus $^{162}$Dy using the SMMC method and demonstrated its rotational character through the low-temperature behavior of $\langle \mathbf{J}^2 \rangle_T$ and the thermal energy $E(T)$. Figure~\ref{low-T} shows these observables as a function of temperature. For comparison we also show for $\langle \mathbf{J}^2 \rangle_T$  the fits to the rotational (solid line) and vibrational (dotted-dashed line) models in Eq.~(\ref{J2-theory}). The SMMC results clearly indicate the rotational character of $^{162}$Dy in the truncated shell model space. Furthermore,  the fit to the rotational model gives a ground-state moment of inertia of $I_{\rm g.s.} =35.5 \pm 3.3$ MeV$^{-1}$, in agreement with the experimental value of $I_{g.s.} = 3/E_{2^+}= 37.2$ MeV$^{-1}$. The solid line in panel (b) is a fit to the rotational model prediction for the thermal energy $E(T)=E_0 +T$, where $E_0$ is the ground-state energy.

\subsection{Crossover from vibrational to rotational collectivity}

\begin{figure}[h]
    \includegraphics[clip,angle=0,width=\columnwidth]{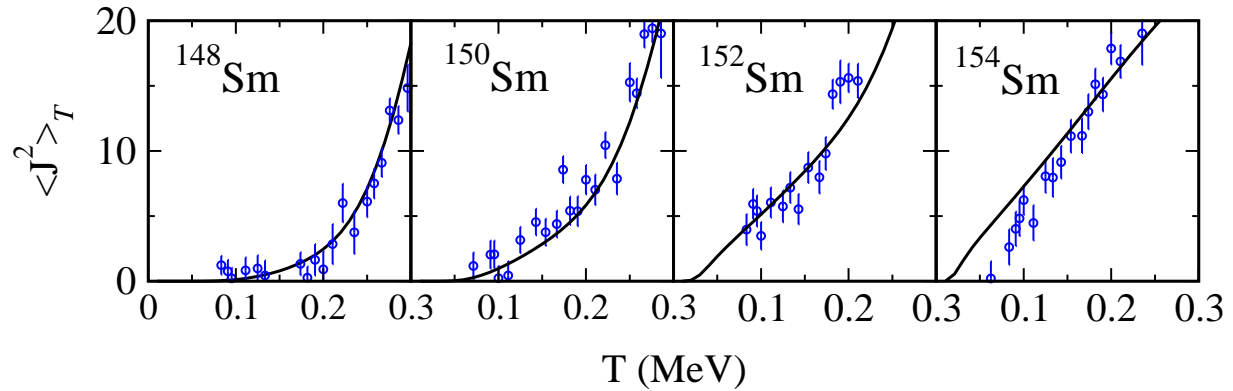}
    \caption{$\langle \mathbf{J}^2 \rangle_T$ as a function of temperature in the even  $^{148-154}$Sm isotopes. The SMMC results (open circles with error bars) are compared with Eq.~(\ref{Eq:J2high}) using a complete set of experimentally known discrete nuclear levels and an experimentally determined BBF state density (except for $^{154}$Sm, for which only discrete nuclear levels are used). Adapted from Ref.~\cite{Ozen2012}. }
    \label{Fig:Sm-J2}
\end{figure}
Even samarium and neodymium isotopes display a shape phase transition from spherical to deformed nuclei as the number of neutrons increases from shell closure towards the mid-shell region. In the SMMC approach, a signature of this transition can be observed in the low-temperature behavior of $\langle \mathbf{J}^2 \rangle_T$. In Fig.~\ref{Fig:Sm-J2} we compare in the even $^{148-154}$Sm isotopes the SMMC results (circles) for $\langle \mathbf{J}^2 \rangle_T$ with results that are derived from the experimental data (solid lines) according to
\begin{eqnarray}
\label{Eq:J2high}
 \langle \mathbf{J}^2 \rangle_T  =  \frac{1}{Z(T)} \left(\sum_i^N J_i(J_i+1)(2J_i+1)e^{-E_{i}/T} +  \right.
    \left. \int_{E_{N}}^\infty d E_x \: \rho(E_x) \: \langle \mathbf{J}^2 \rangle_{E_x} \; e^{-E_x/T} \right)\;.
\end{eqnarray}
Here $Z(T)=\sum_{i}^{N} (2J_i+1) e^{-E_i/T} + \int_{E_{N}}^\infty d E_x \rho(E_x) e^{-E_x/T}$ is the corresponding experimental partition function. The summations in (\ref{Eq:J2high}) and in the expression for $Z(T)$ are carried over a complete set of experimentally known low-lying levels $i$ with excitation energy $E_i$  and spin $J_i$, and $\rho(E_x)$ is the back-shifted Bethe formula (BBF) with parameters that are determined from level counting and neutron resonance data.  $\langle \mathbf{J}^2 \rangle_{E_x}$ (the average value of $\mathbf{J}^2$ at a given excitation energy $E_x$) is calculated from $\langle \mathbf{J}^2 \rangle_{E_x}=3\sigma^2$, where $\sigma^2$ is the rigid-body spin-cutoff parameter.  We find good agreement between the SMMC results and the experimental estimates. $^{148}$Sm is spherical in its ground state and its $\langle \mathbf{J}^2 \rangle_T$  exhibits a ``soft'' response to temperature, typical of vibrational nuclei.
In contrast, $\langle \mathbf{J}^2 \rangle_T$ in  $^{154}$Sm exhibits a linear response to temperature, indicating the rotational character of this nucleus.  A similar
study of even $^{144-152}$Nd isotopes~\cite{Ozen2012} confirms that  $^{144}$Nd and $^{146}$Nd are vibrational, while $^{150}$Nd and  $^{152}$Nd are rotational.

\section{State densities in heavy nuclei}\label{density}

\subsection{State densities}
\begin{figure}[tbh]
    \includegraphics[clip,angle=0,width=\columnwidth]{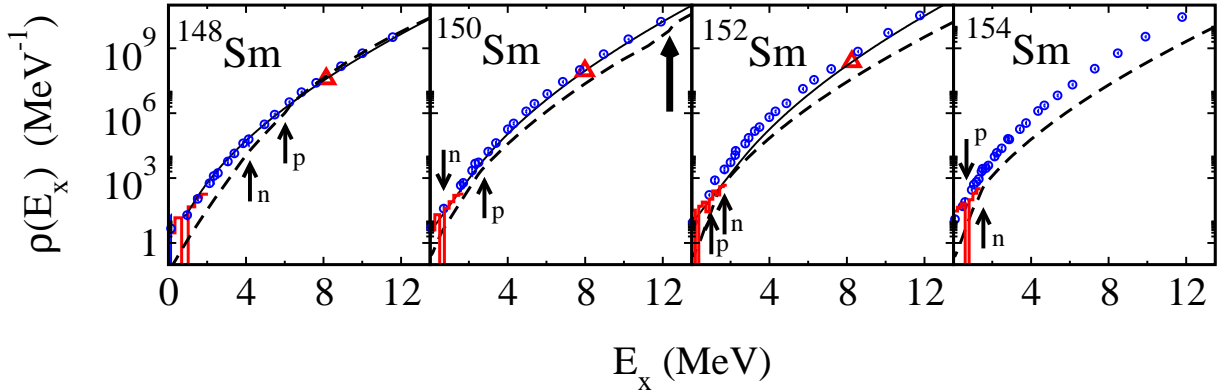}
    \caption{Total state densities of the even $^{148-154}$Sm isotopes. The SMMC results (open circles) are
    compared with the level counting data (histograms), the neutron resonance data (triangles), the experimentally parametrized BBF
    state densities (solid lines)
    and the HFB densities (dashed lines). Arrows denote proton and neutron pairing phase transitions and the thick arrow denotes a shape transition. Adapted from Ref.~\cite{Ozen2012}.}
    \label{Fig:Sm-rho}
\end{figure}
The SMMC method has been successful in the calculation of state densities in medium-mass nuclei~\cite{Nakada1997,SMMC-statistical}. In the SMMC method we calculate the average thermal energy $E(\beta)$ as a function of $\beta$, from which we can compute the canonical partition function. The average state density is then determined by applying the saddle-point approximation to the integral that expresses the state density  as the inverse Laplace transform of the canonical partition function.  

In Fig.~\ref{Fig:Sm-rho}  we show the SMMC state densities (open circles) in the even $^{148-154}$Sm isotopes and compare them with level counting data at low energies (histograms) and neutron resonance data (triangles) when available. The solid lines are BBF state densities that are determined from fits to experimental data. The dashed lines are the densities obtained in the finite-temperature Hartree-Fock-Bogoliubov (HFB) approximation. The HFB results describe the intrinsic density of states, and the enhancement observed in the SMMC state density arises from collective states (vibrational and rotational) that are built on top of the intrinsic states. In the HFB approximation we observe proton and neutron pairing phase transitions (arrows) and shape phase transitions (thick arrows). $^{148}$Sm is spherical in its ground state and exhibits pairing transitions only, while $^{150}$Sm is deformed in its ground state and thus undergoes also a shape transition to a spherical shape at $E_x \approx 12.5$ MeV.  Shape phase transitions also occur in $^{152}$Sm and  $^{154}$Sm at excitation energies that increase with mass number (not shown in the figure).

\subsection{Collective enhancement factors}
\begin{figure}[tbh]
    \includegraphics[clip,angle=0,width=\columnwidth]{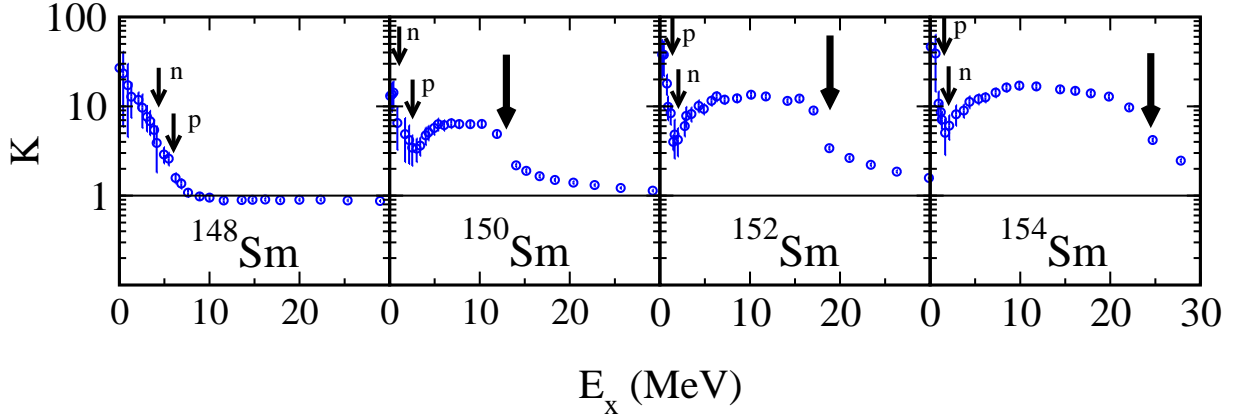}
    \caption{Total collective enhancement factor $K$ (see text) in the even $^{148-154}$Sm isotopes as a function of excitation energy $E_x$. Arrows are as in Fig.~\ref{Fig:Sm-rho}. Adapted from Ref.~\cite{Ozen2012}.}
    \label{Fig:Sm-enh}
\end{figure}
Collective effects in the nuclear state density are difficult to model microscopically  and are usually accounted for by phenomenological vibrational and rotational collective enhancement factors~\cite{RIPL}.  In Ref.~\cite{Ozen2012} we have proposed to use
the ratio of the SMMC and the HFB state densities, $K = \rho_{\rm SMMC}/\rho_{\rm HFB}$, as a microscopic measure of the collective enhancement factor. In Fig.~\ref{Fig:Sm-enh}  we show $K$ as a function of the excitation energy $E_x$ for the same samarium isotopes shown in Fig.~\ref{Fig:Sm-rho}.  In $^{148}$Sm, which is spherical in
its ground state, any collectivity should be exclusively due to vibrational excitations, whereas the heavier deformed samarium nuclei are expected to have both vibrational and rotational collectivity. We observe that all collectivity disappears
in $^{148}$Sm (i.e., $K \approx 1$) at excitation energies above the pairing transitions,  suggesting  that the decay of the vibrational collectivity is associated with the pairing phase transitions. In the heavier samarium isotopes $K$ reaches a local minimum above the pairing transition energies but increases at higher energies until it reaches a plateau and then decays to unity in the vicinity of the shape transition energy. This behavior can be explained by the contribution of rotational collectivity to $K$ that decays to unity once the nucleus becomes spherical and can no longer support rotational bands.

\section{Odd-even and odd-odd nuclei}
The projection on an odd number of particles leads to a sign problem in SMMC  (even for good sign interactions). This problem is particularly severe at low temperatures and makes it difficult to calculate the ground-state energy of odd-even and odd-odd nuclei. We have developed a method that circumvents this sign problem for the ground-state energy  by using asymptotic properties of the single-particle Green's functions of the neighboring even-even nuclei~\cite{Mukherjee2012}.

\subsection{Green's function method}\label{Green}
For a rotationally invariant Hamiltonian, the scalar imaginary-time single-particle
Green's functions are given by
\be\label{green}
G_{\nu}(\tau) = \frac{\rTr_{\ma}\left[~e^{-\beta H} \mathcal{T} \sum_m a_{\nu m}(\tau) \adag_{\nu m}(0)\right]}{ \rTr_{\ma}~e^{-\beta H}}\;,
\ee
where $\nu \equiv (n l j)$ labels the nucleon single-particle orbital with radial quantum number $n$, orbital angular momentum $l$ and total spin $j$. Here $\mathcal{T}$ denotes
time ordering and $a_{\nu m}(\tau)\equiv e^{\tau H} a_{\nu m} e^{-\tau H}$ is an annihilation operator of a nucleon at imaginary time $\tau$ ($-\beta \leq \tau\leq \beta$) in a single-particle state with orbital $\nu$ and magnetic quantum number $m$.

Consider an even-even nucleus $\ma \equiv (Z,N)$, with $Z$ protons and $N$ neutrons. Assuming that the ground state of this nucleus has spin 0, the $\tau$-dependence of the Green's function has the asymptotic form  $G_{\nu}(\tau) \sim e^{- \Delta E_{J=j}(\ma_{\pm}) |\tau|}$, where $\ma_\pm$ denote the even-odd nuclei $(Z,N \pm 1)$ when $\nu$ is a neutron orbital and the odd-even nuclei $(Z \pm 1,N)$ when $\nu$ is a proton orbital.  The $+$ and $-$ subscripts should be used for $\tau > 0$ and $\tau \leq 0$, respectively. $\Delta E_{J=j}(\ma_{\pm})$ is the difference between the energies of the lowest spin $J$ eigenstate of the $\ma_{\pm}$-particle nucleus and the ground state of the $\ma$-particle nucleus.  In this asymptotic regime for $\tau$, we can calculate $\Delta E_j(\ma_{\pm})$ from the slope of $\ln G_{\nu}(\tau)$. The minimum of $\Delta E_j(\ma_{\pm})$ over all possible values of $j$ gives the difference between the ground-state energy of the $\ma_{\pm}$ nuclei, $E_{\rm gs}(\ma_{\pm})$,
and the ground-state energy of the $\ma$ nucleus, $E_{\rm gs}(\ma)$. Both $E_{\rm gs}(\ma)$ and $G_{\nu}(\tau)$ are related to the even-even nucleus and can be calculated in SMMC without a sign problem (for a good sign interaction).

\subsection{Pairing gaps in iron-region nuclei}
\begin{figure}[h]
\includegraphics[width= 0.85\textwidth]{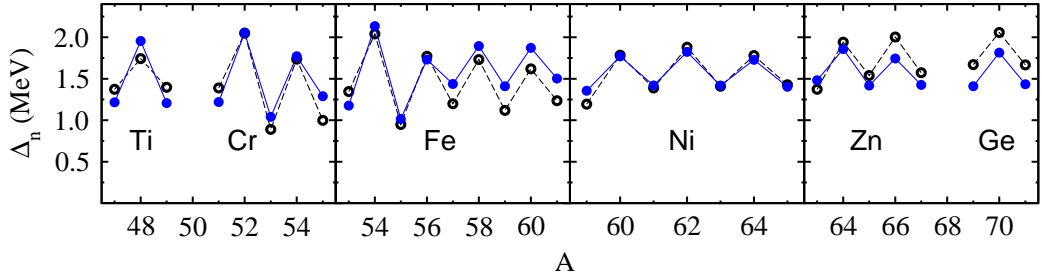}\centering
\caption{\label{fig2} Neutron pairing gaps $\Delta_n$ as a function of mass number $A$ in $fpg_{9/2}$-shell nuclei. The gaps calculated with the Green's
function method (solid circles) are compared with the experimental gaps (open circles).
Adapted from Ref.~\cite{Mukherjee2012}.}
\end{figure}
The method of Sec.~\ref{Green} is particularly useful for calculating pairing gaps (i.e., odd-even staggering of masses) because the odd-even ground state energy differences are calculated directly. In Fig.~\ref{fig2} we compare the SMMC pairing gaps of iron-region nuclei with their experimental values. We have used the complete $fpg_9/2$-shell model space with the Hamiltonian of Ref.~\cite{Nakada1997}. The statistical errors for the SMMC pairing gaps ($\sim 0.01$ MeV) are not visible in the figure. The theoretical gaps are in reasonable agreement with the experimental values.

\section{Conclusion}
We have discussed recent progress in the SMMC approach. We have shown that the crossover from vibrational to rotational collectivity in rare-earth nuclei can be described microscopically in the framework of the CI shell model approach. We have also described a method to calculate the ground-state energy of odd-even and odd-odd nuclei that circumvents the odd-particle sign problem and applied it to calculate pairing gaps of nuclei in the iron region.

\ack

It is a great pleasure for us to dedicate this article to Jerry Draayer on the occasion of his 70th birthday. This work was supported in part by the DOE grant DE-FG-0291-ER-40608, and by the JSPS Grant-in-Aid for Scientific Research (C) No.~22540266. Computational cycles were provided by the NERSC and  Yale University High Performance Computing Center.

\end{document}